\documentclass{note_mod}
  \pdfoutput=1
\usepackage{atlasphysics}
\usepackage{subfigure}
\usepackage{hyperref}
\usepackage{authblk,ifthen}
\usepackage{xspace}
\usepackage{multirow}
\usepackage{lineno}
\usepackage{amssymb,amsmath}
\usepackage{lscape}
\usepackage{caption}
\usepackage{enumerate}
\usepackage{textcomp}
\usepackage{array}
\usepackage{verbatim}
\skipbeforetitle{100pt}

\title{Heavy quarks and quarkonia production in high-energy experiments}

\author[]  {Sandro Palestini} 

\affil[]{CERN, CH 1211 Geneva 23, Switzerland}

\journal{XVII International Conference on Hadron Spectroscopy and Structure, \\ Hadron2017, 25-29 September, 2017, University of Salamanca, Salamanca, Spain.} 

\abstracttext{
\noindent 
Proton-proton collisions at the LHC provide the ground for tests of the strong interactions 
through the study of the production mechanisms for quarkonia and hadrons containing charm or beauty flavours.
This review addresses recent results and includes 
studies of kinematical correlations, polarisation and associated production.
QCD-based models are briefly discussed and compared to measurements. 
}

\begin{document}

\section{Introduction}
The remarkable magnitudes of collision energy and integrated luminosity achieved in proton-proton collisions
at the LHC have opened the possibility of detailed tests of QCD through studies of the production of hadron containing heavy quarks and of quarkonia.
After brief introduction to the theory framework, the following sections discuss results on the production of $c$ and $b$ hadrons,
on $b$--$\overline b$ kinematical correlations, on quarkonia  ($c\overline c$ and $b\overline b$) production, on quarkonia polarisation,
and on the associated production of $J\!/\!\psi$ pairs. 

\section{Production of hadrons containing heavy quarks in hadron collisions} 
The formal description of the production cross-section of a hadron $H_Q$ ($Q = c,\, b$) in high-energy hadron collisions is based on the factorisation formula
\begin{equation}
\frac{d \sigma^Q}{dX}=\sum_{i,j} \int \! d\hat {\vphantom{\scalebox{1.15}{X}}X} \int \! dx_1 \int \! dx_2 \, f_i(x_1,q)
\, f_j{}^\prime(x_2,q) \frac{d\hat{\vphantom{b}\sigma}{}^Q_{i,j}(x_1,x_2,s,q)}{d\hat{\vphantom{\scalebox{1.15}{X}}X} }
\, F(\hat {\vphantom{\scalebox{1.15}{X}}X} ,X,q)\: ,  \label{eq:factorisation}
\end{equation}
where the final states is specified by the set of variables $X$, 
and the process is described as the strong-interaction scattering of partons $i, j$,
with Feynman variables $x_1$, $x_2$,  into the state $\hat {\vphantom{\scalebox{1.15}{X}}X} $, 
integrated over the parton distributions (PDFs) $f_i, f_j{}^\prime$ of the colliding hadrons, 
with the fragmentation function $F$ describing the transition 
to the final state, particle-level variables.   
The PDF's are universal and are extracted from fits to processes where  
the parton scattering is known with high accuracy. 
Similarly, the function $F$ includes long-distance effects, and can be extracted from independent observations ({\em e.g.}\ $b$-hadron fragmentation in $e^+\!e^-\! \rightarrow Z \rightarrow b \, \overline b$). 

All terms in the factorisation equation depend logarithmically on energy scales, represented by $q$ in 
equation~\ref{eq:factorisation}.  Besides the factorisation scale, which affects the PDF's, and the renormalisation scale of the strong coupling $\alpha_s$, which enters the parton cross-section, the production of heavy quarks depends also on the scales set by the heavy quark mass ($m_Q$) and by the transverse momentum ($p_{\mathrm T}$) of the hadron $H_Q$.

Different ranges of $p_{\mathrm T}$ imply different computational methods. For values 
$p_{\mathrm T} \lesssim 5\times  m_Q$,  
only light flavours are included as active constituents ($u, d, s$ and also $c$ for beauty production, together with 
the gluon $g$)  in the Fixed-Flavour-Number-Scheme (FFNS).  
For larger values of $p_{\mathrm T}$, the logarithm  $\ln (p_{\mathrm T} /m_Q)$ is large and requires resummation, 
the heavy quarks are treated as active partons (Variable-Flavour-Number-Scheme VFNS),
and all quarks may be treated as massless (Zero-Mass VFNS). 
In the fragmentation function $F$, a scale-dependent pertubative term may be combined with the scale independent term determined from fits to data.

Unified frameworks merging the low $p_{\mathrm T}$, massive computation and the high $p_{\mathrm T}$, massless computation have been developed, in particular the General-Mass VFNS (GM-VFNS)~\cite{ref:GM-VFNS} and the 
Fixed-Order plus Next-to-Leading-Logarithms (FONLL)~\cite{ref:FONLL} schemes.  Both methods rely on perturbative computations performed at Next-to-Leading-Order (NLO) and Next-to-Leading-Logarithms (NLL). 
Besides uncertainties related to the limited perturbative sums, the validities of these computation is related to assumptions in the transitions between different regimes, and in the use of non-perturbative fragmentation functions. 
Recent experimental results, discussed below, have extended the test of the prediction to larger ranges 
in $p_{\mathrm T}$ and rapidity $y$, and to higher collision energies $\sqrt s$. 

The experimental observables are 
frequently compared to Monte Carlo generators designed to model the production mechanism and the hadronisation process. PYTHIA~\cite{ref:PYTHIA}  and HERWIG~\cite{ref:HERWIG} describe the parton scattering at leading order (LO) and leading logarithm (LL) approximation, but NLO modelling has become available with the development of MC@NLO~\cite{ref:MC@NLO}, POWHEG~\cite{ref:POWHEG} and MADGRAPH~\cite{ref:MADGRAPH}.


\subsection{Measurements of charm and beauty production}  \label{sec:CBprod} 
\begin{figure}[!tb]
\begin{center}
   \hspace{-1cm}
   \begin{minipage}{0.45\textwidth}
   \begin{center}
   \includegraphics[width=0.97\textwidth]{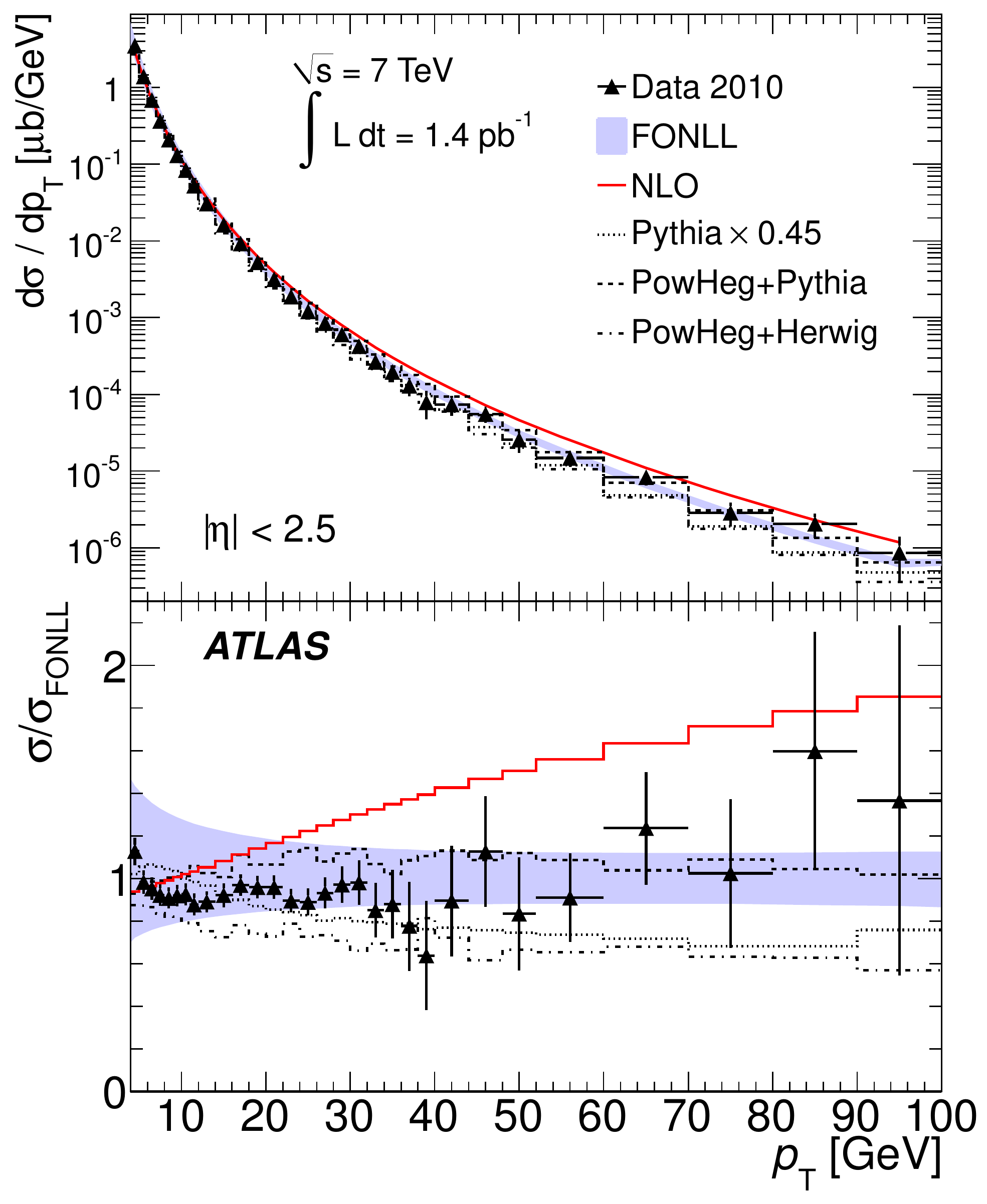} 
   \caption{Muon differential cross-section for charm and beauty semileptonic decays, vs. muon $p_{\mathrm T}$. 
   From reference~\cite{ref:lepton-spectrumATLAS}.}
     \label{fig:mu-spectrum}
     \end{center}
    \end{minipage}%
   \hspace{0.5cm}
   \begin{minipage}{0.50\textwidth}
 \includegraphics[width=1.1\textwidth]{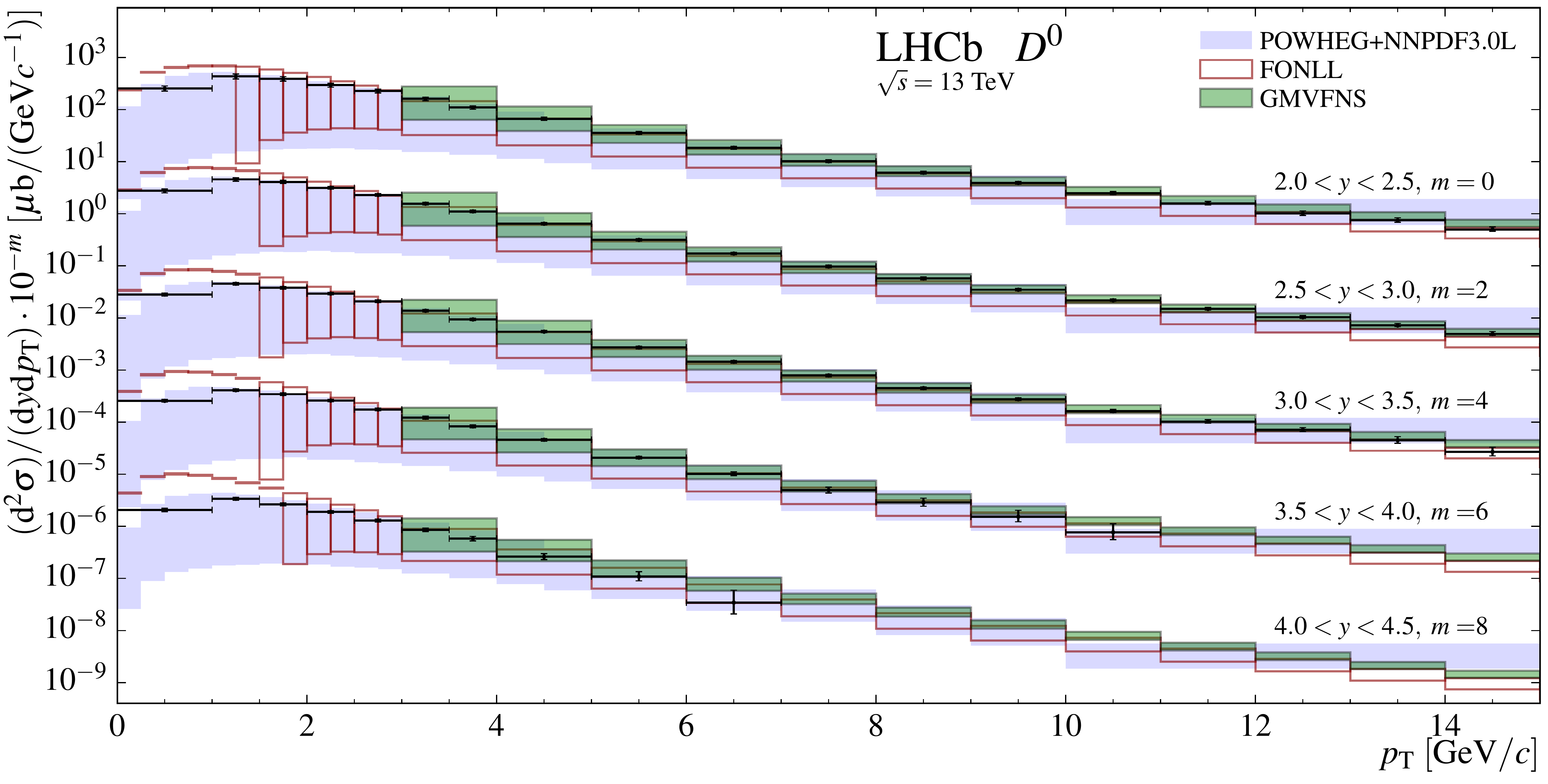} \\
 \includegraphics[width=1.1\textwidth]{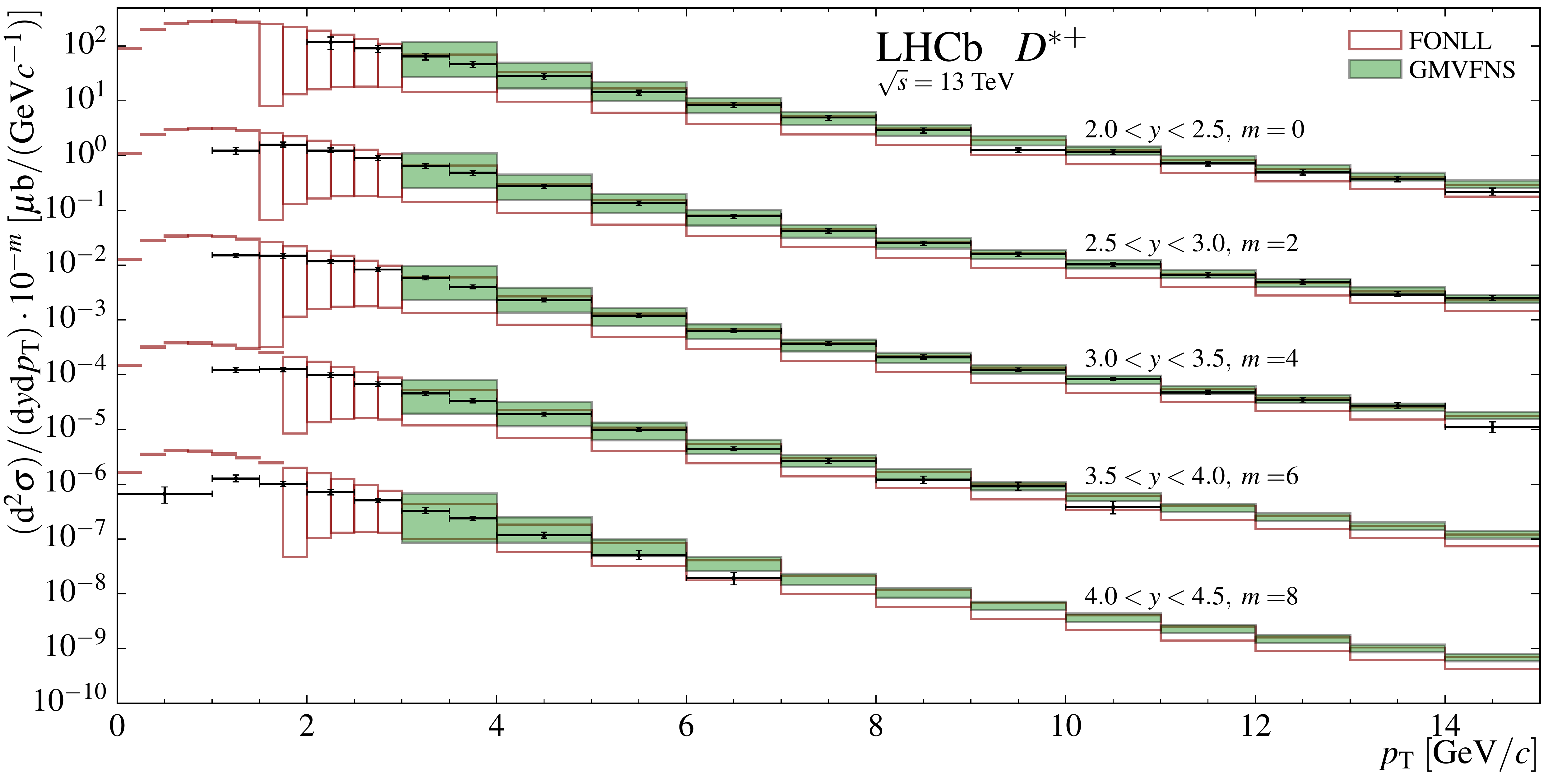} 
  \begin{center}
    \caption{$D^{\,o}$, $D^{*+}$ production in the forward region. From reference~\cite{ref:charm_LHCb}.} 
    \label{fig:D-prod-LHCb}
 \end{center}
    \end{minipage}
  \end{center}
\end{figure}

The production of 
hadrons containing heavy quarks 
may be studied using fully reconstructed final states, or  
with partially reconstructed final states. An example of the latter approach is shown in figure~\ref{fig:mu-spectrum}, where the $p_{\mathrm T}$ spectrum of muons for pseudo-rapidity $|\eta | <$ 2.5 is provided by ATLAS~\cite{ref:lepton-spectrumATLAS} after subtracting the contribution from Drell-Yan continuum and $W$, $Z$ decays. 
Semileptonic decays of charm and beauty hadrons contribute to the spectrum.
The data are compared to predictions 
and found in reasonable agreement, in particular with FONLL. The spectrum indicated with NLO shows the effect of removing the NLL component in FONLL.

LHCb has studied beauty and charm production in the forward, low $p_{\mathrm T}$  region. 
Figure~\ref{fig:D-prod-LHCb} shows the measurement of $D^{\,o}$ and $D^{*+}$ mesons, with $D^+$ and $D^+_s$ and  also presented in reference~\cite{ref:charm_LHCb}. While usually the measurement of heavy-quark production cross-section is used to test the short-distance parton scattering and not the parton PDFs, the ratio between the measurements at 13 and 7~TeV in the forward region 
reaches the sensitivity for a test of the gluon distribution function~\cite{ref:gluonPDF}.

\begin{figure}[!t]
  \begin{center}
  \hspace{-0.3cm}
    \includegraphics[width=0.5\textwidth]{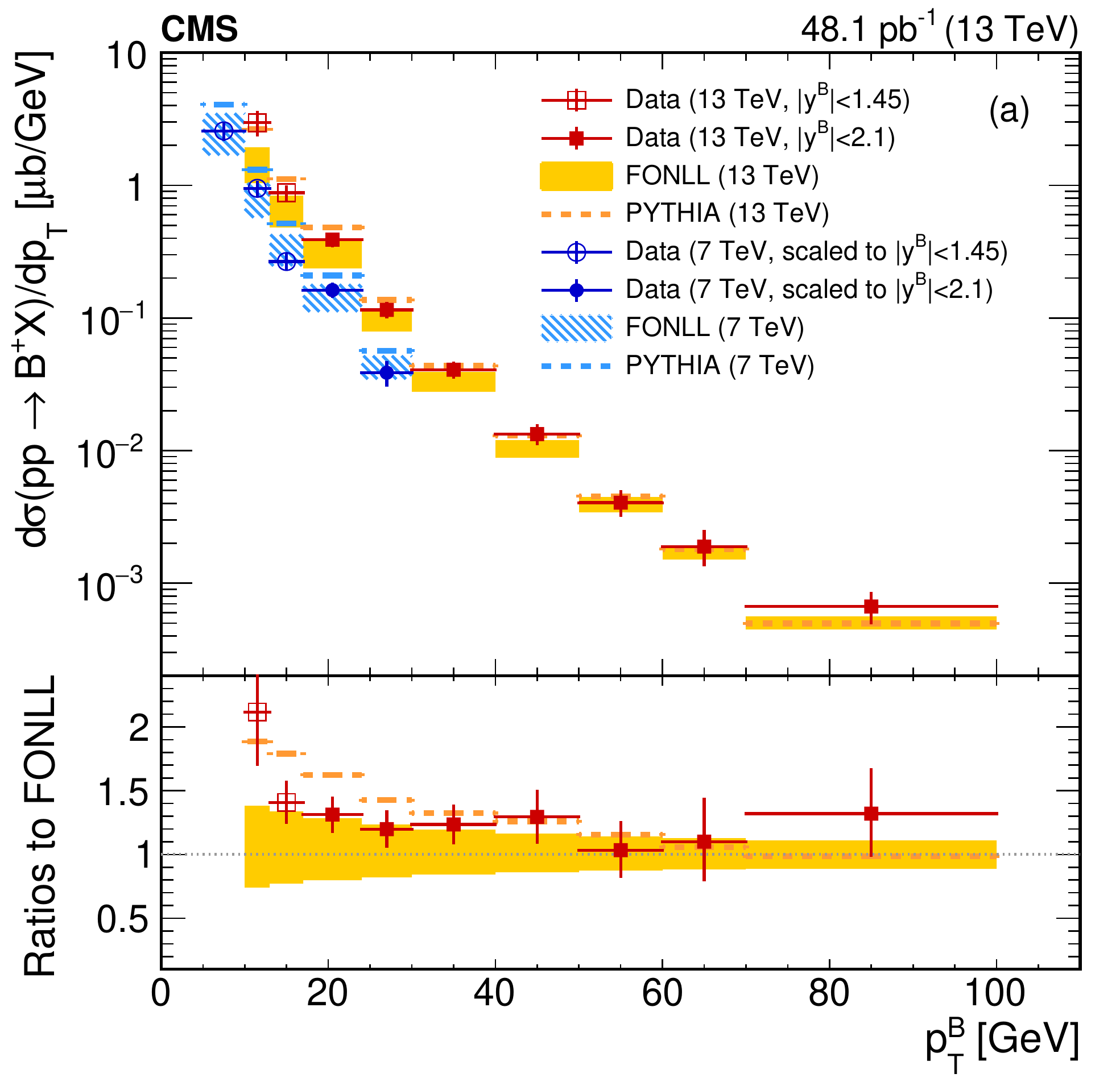} 
  \hspace{+0.1cm}
    \includegraphics[width=0.25\textwidth]{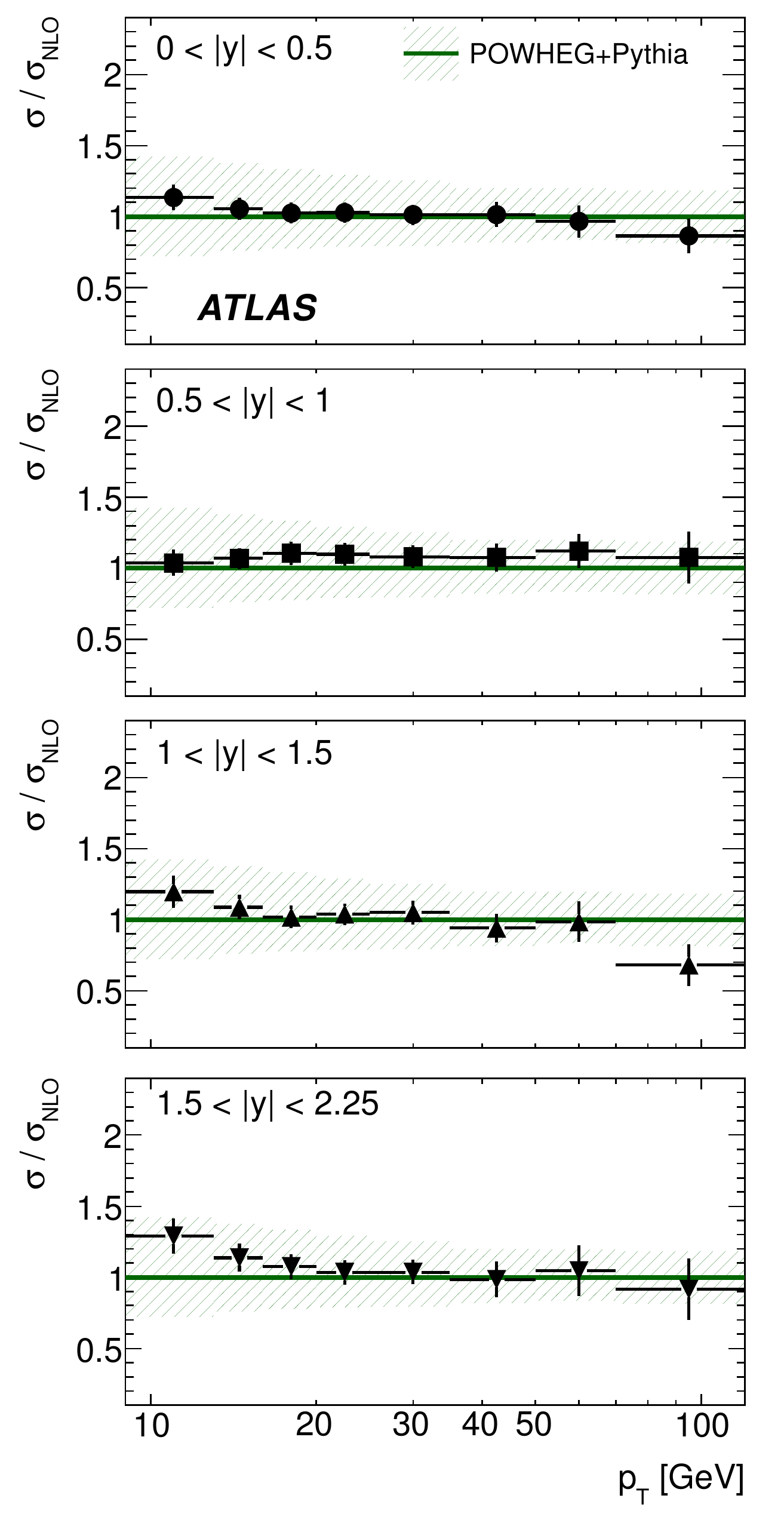} 
  \hspace{-0.3cm}
      \includegraphics[width=0.25\textwidth]{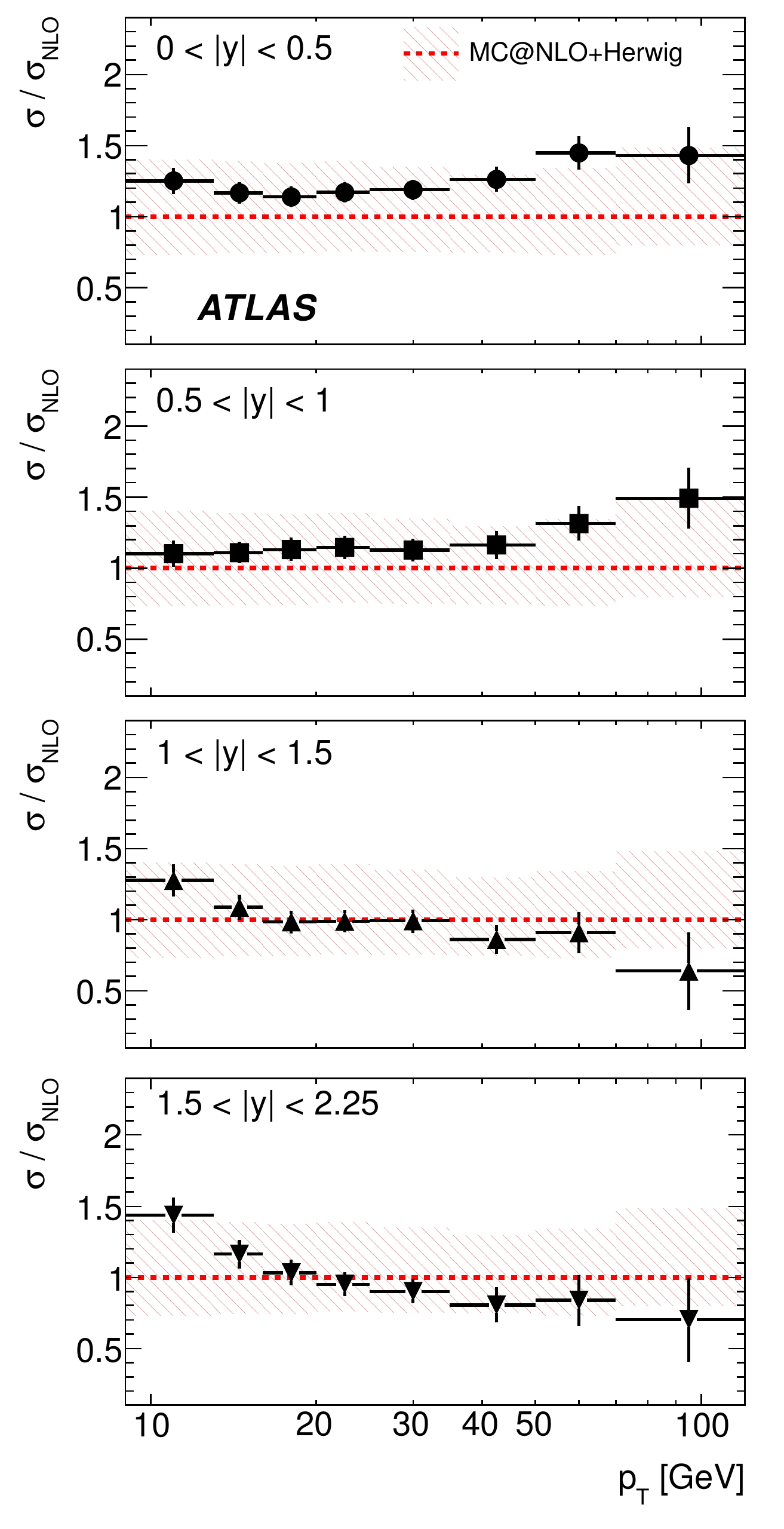} 
    \caption{$B^+$ differential production cross-section as a function of $p_{\mathrm T}$ in the central rapidity region, compared with NLO computations, at 7 and 13~TeV ({\em left}), and for different $y$ intervals at 7~TeV ({\em right}).  From references~\cite{ref:BprodCMS} and \cite{ref:BprodATLAS}.} 
    \label{fig:B+prod-CMS-ATLAS}
  \end{center}
\end{figure}

\begin{figure}[!b]
  \begin{center}
  \hspace{-0.2cm}
    \includegraphics[width=0.5\textwidth]{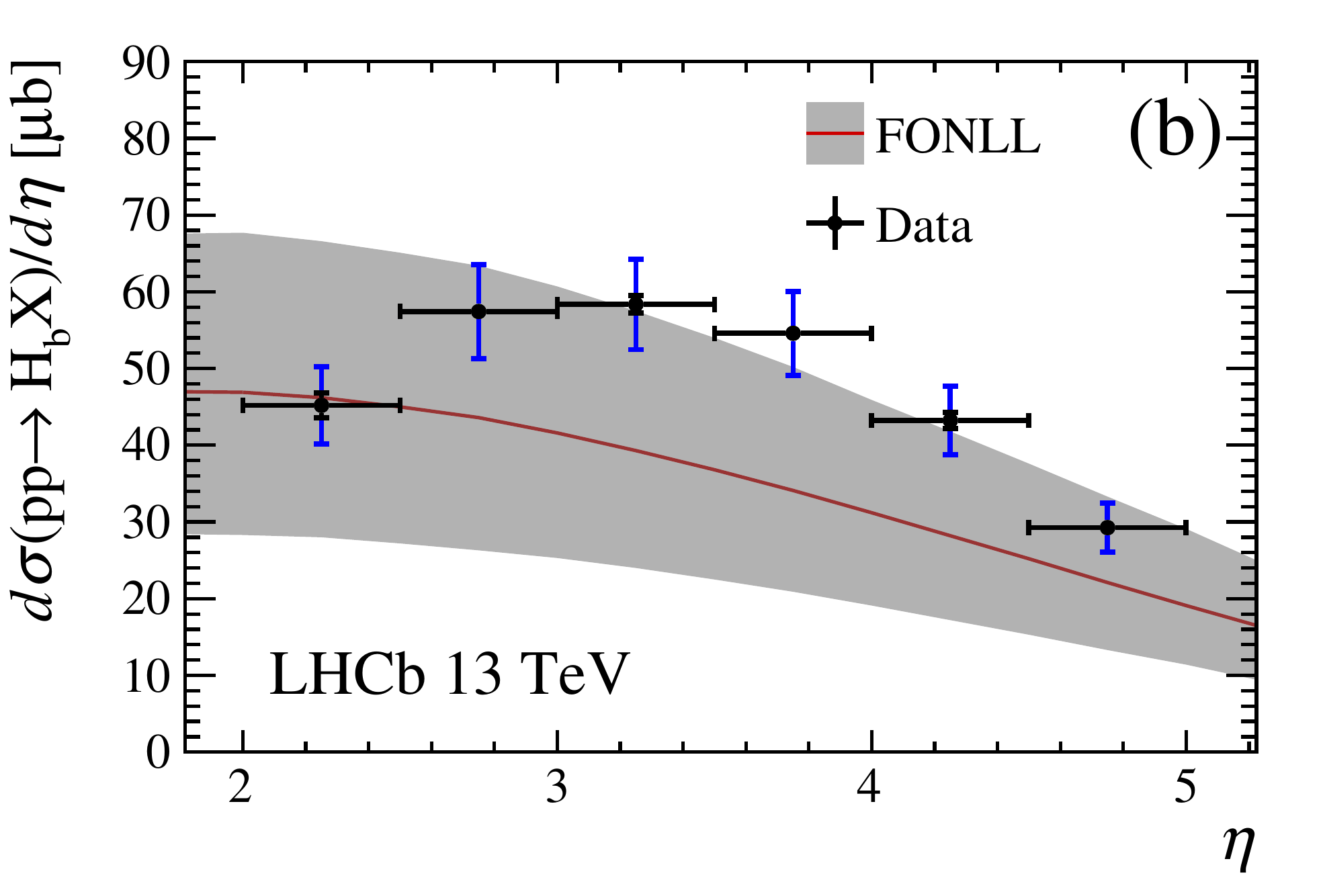} 
      \includegraphics[width=0.5\textwidth]{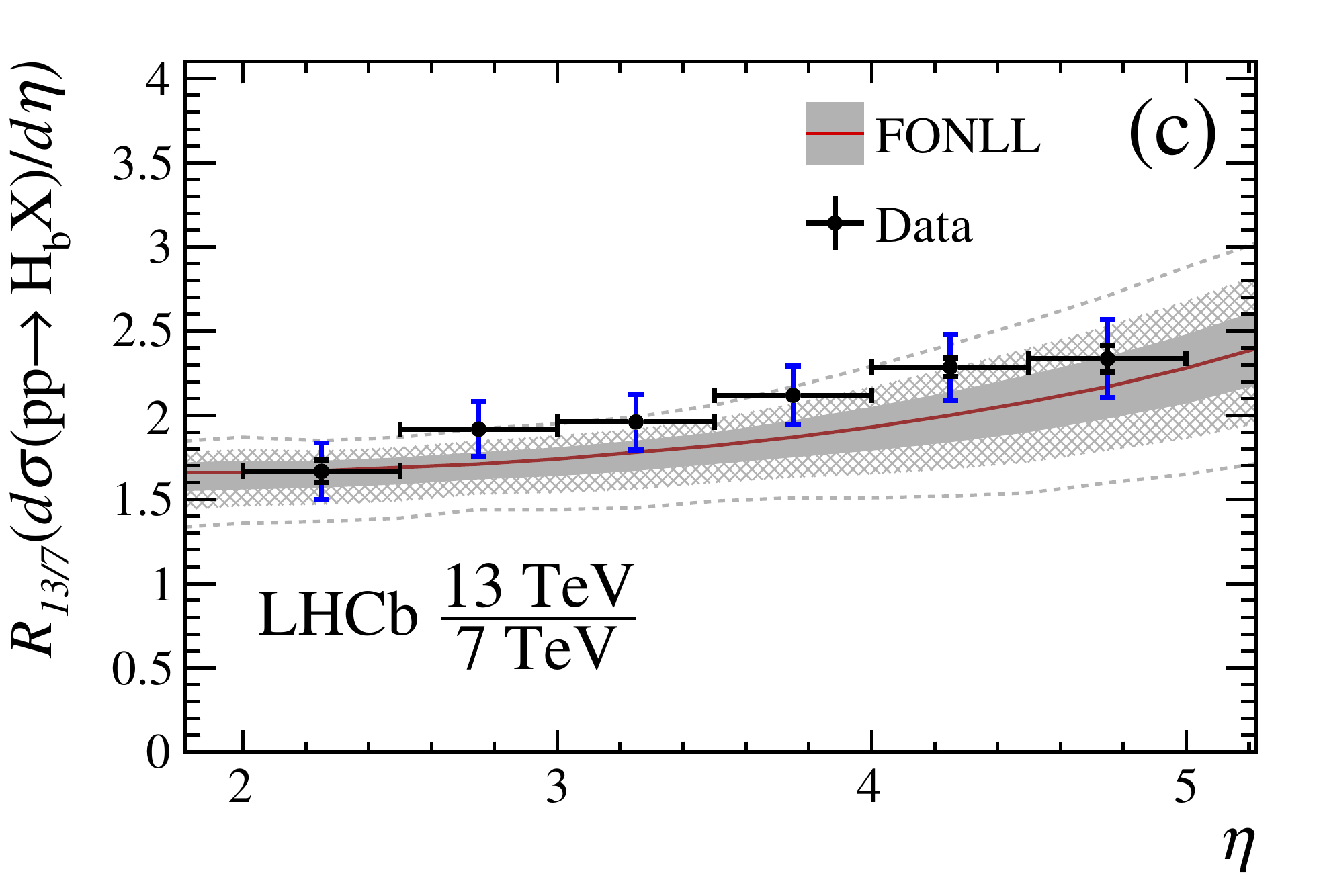} 
    \caption{Rapidity dependence of the $b$-hadron production cross-section 
    and of the ratio of the cross-section at 13 and 7~TeV, in the forward region.  From reference~\cite{ref:BprodLHCb}.} 
    \label{fig:B-prod-LHCb}
  \end{center}
\end{figure}

The production cross-section of $B^+$ has been measured in the central rapidity region ($|y| \lesssim 2.2$) 
by CMS~\cite{ref:BprodCMS} and ATLAS~\cite{ref:BprodATLAS}
in the decay to $J\!/\!\psi \, K^+  \rightarrow \mu^+ \mu^- K^+$, with the muon pair used for on-line selection.
Figure~\ref{fig:B+prod-CMS-ATLAS} shows good agreement with FONLL and POWHEG for $p_{\mathrm T} > 15$~GeV. 
Figure~\ref{fig:B-prod-LHCb} shows the measurement of LHCb in the forward region~\cite{ref:BprodLHCb}, 
where the dependence on the rapidity is only in fair agreement with FONLL, which instead describes nicely the increase in production rate 
between 7 and 13~TeV. 
Production cross-sections have been measured also for $\Lambda^o_b$~\cite{ref:L_CMS, ref:L_B0_LHCb},
$B^{\, o}$~\cite{ref:L_B0_LHCb}, and for the ratio of production cross-section times branching-fraction to final state 
between $B^+_c\rightarrow J\!/\!\psi\, \pi^+$ and $B^+\rightarrow J\!/\!\psi\, K^+$~\cite{ref:Bc_CMS,ref:Bc_LHCb}.

\subsection{Correlations in $b $-$ {\overline b}$ production} \label{sec:Bcorr}
\begin{figure}[!tb]
  \begin{center}
  \hspace{-0.3cm}
    \includegraphics[width=0.5\textwidth]{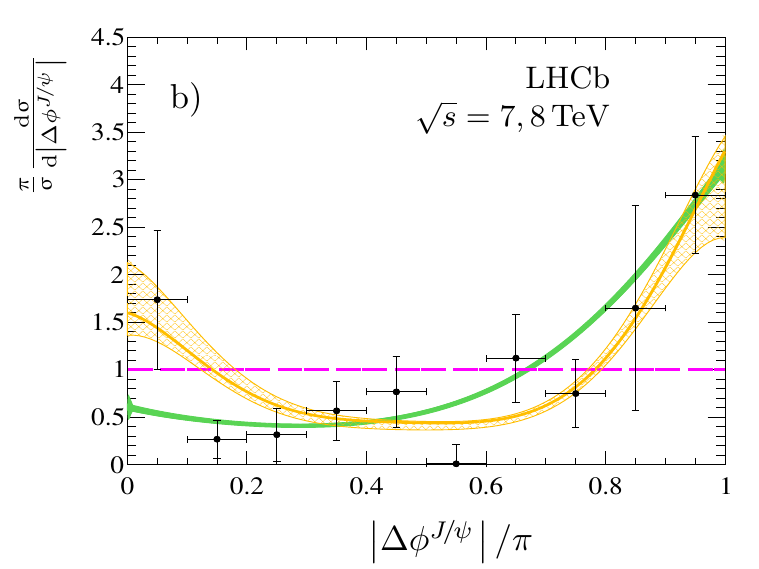} 
      \includegraphics[width=0.5\textwidth]{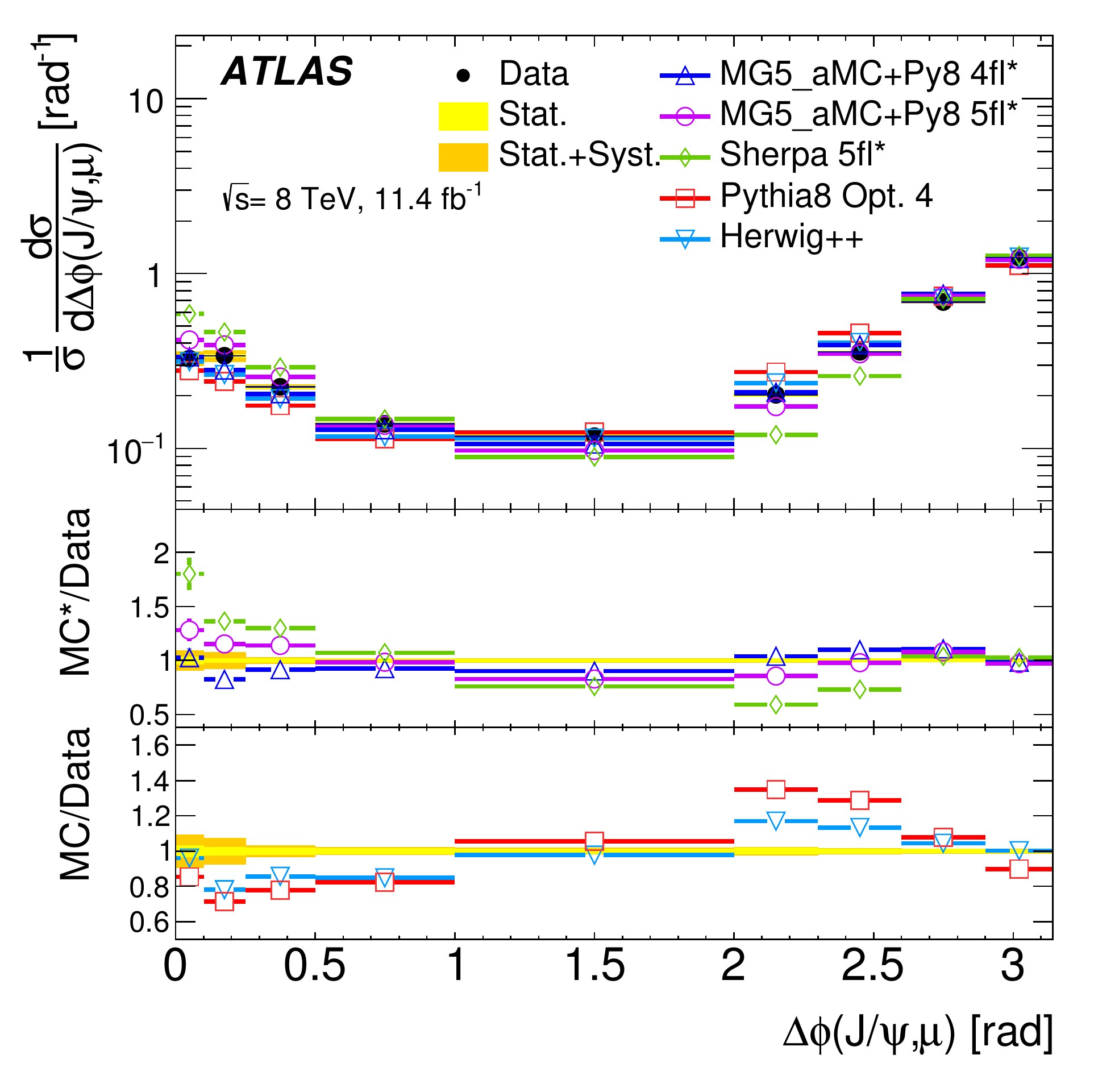} 
    \caption{{\em Left:} azimuthal angular separation between $J\!/\!\psi$'s from $b$, $\overline  b$ in the forward region, with $p_{\mathrm T}{}^{J\!/\!\psi} > 7$~GeV. The yellow (green) band shows the POWHEG (PYTHIA) prediction. 
     From reference~\cite{ref:Bcorre_LHCb}. 
{\em Right:} azimuthal angles separation between $J\!/\!\psi (\mu^+\mu^-)$ and $\mu^\pm$ from $b$, $\overline  b$ in the central region, with $p_{\mathrm T}{}^\mu > 6$~GeV, compared to predictions. 
From reference~\cite{ref:Bcorre_ATLAS}.} 
    \label{fig:BBcorrelation}
  \end{center}
\end{figure}

The kinematical correlations between $b$-  and $\overline b$-hadrons have recently attracted interest.
At the LHC energies, the $t$ channel  $g\,g \rightarrow b\,\overline b$ process dominates the LO contribution,
while at NLO the gluon splitting $g \rightarrow b\,\overline b$ is relevant at high $p_{\mathrm T}$
and for small opening angle between the two hadrons.
Figure~\ref{fig:BBcorrelation}-left shows a recent result from LHCb~\cite{ref:Bcorre_LHCb},
based on pairs of non-prompt $J\!/\!\psi$'s,
with the data favouring the NLO enhancement at small difference in azimuthal angle. 
The expected enhancement is more evident in the data collected by 
ATLAS~\cite{ref:Bcorre_ATLAS}, looking at non-prompt $J\!/\!\psi$ and $\mu^\pm$, 
in a region of higher $p_{\mathrm T}$, as shown in figure~\ref{fig:BBcorrelation}-right. 
NLO computations, in particular MADGRAPH5\_aMC+PYTHIA/4-flavours, provide a rather accurate description of the data. 

\section{Quarkonia production} 
The prompt production of {\em hidden} heavy flavour, {\em i.e.}\ of quarkonia $c\overline c$, $b\overline b$ bound states, differs from the production of {\em open} heavy flavour, since long-distance, non-perturbative effects enter the process more directly than the quasi-universal fragmentation fraction $F$ of equation~\ref{eq:factorisation}. 

\begin{figure}[!t]
  \begin{center}
  \hspace{-0.3cm}
    \includegraphics[width=0.5\textwidth]{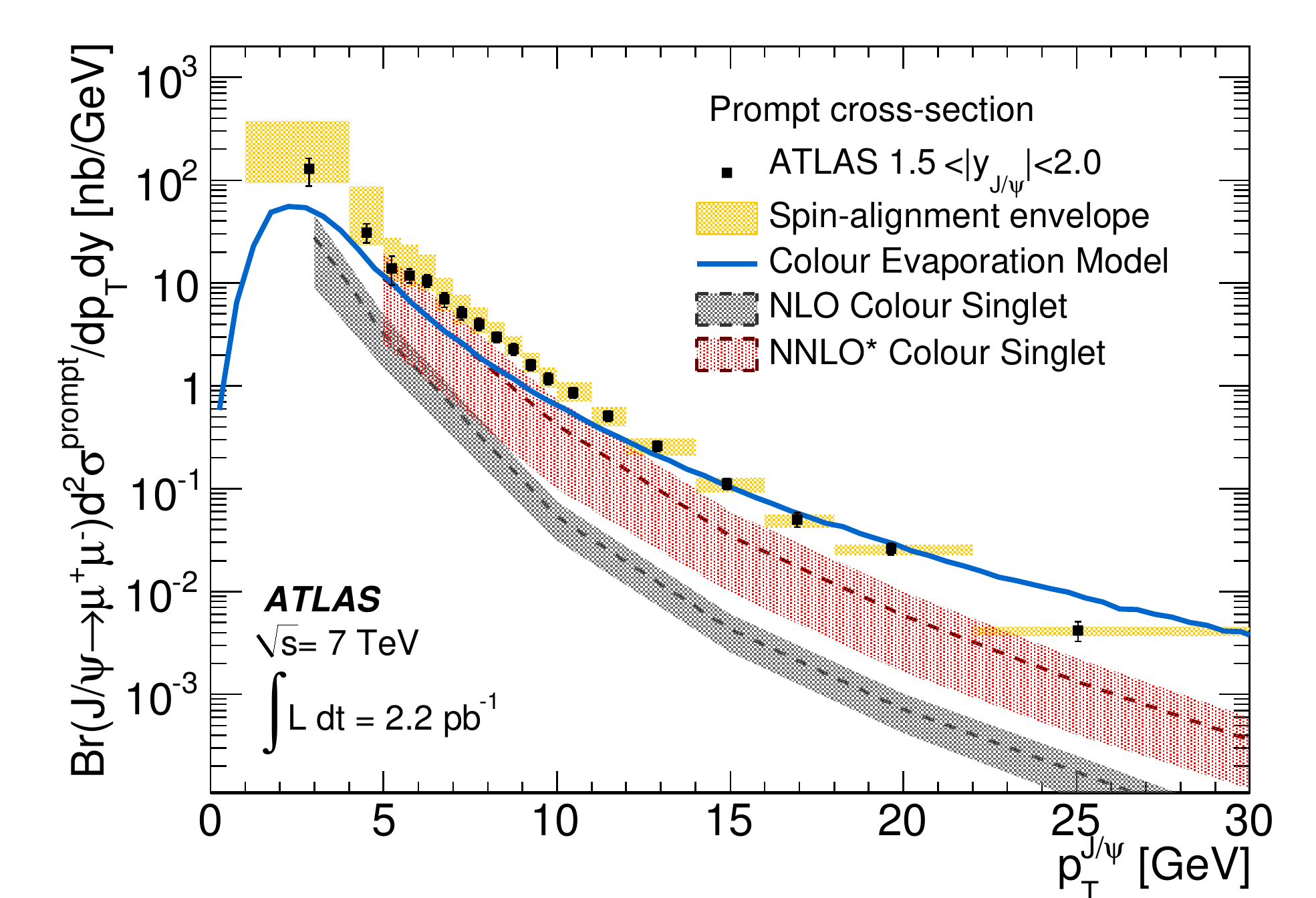} 
      \includegraphics[width=0.5\textwidth]{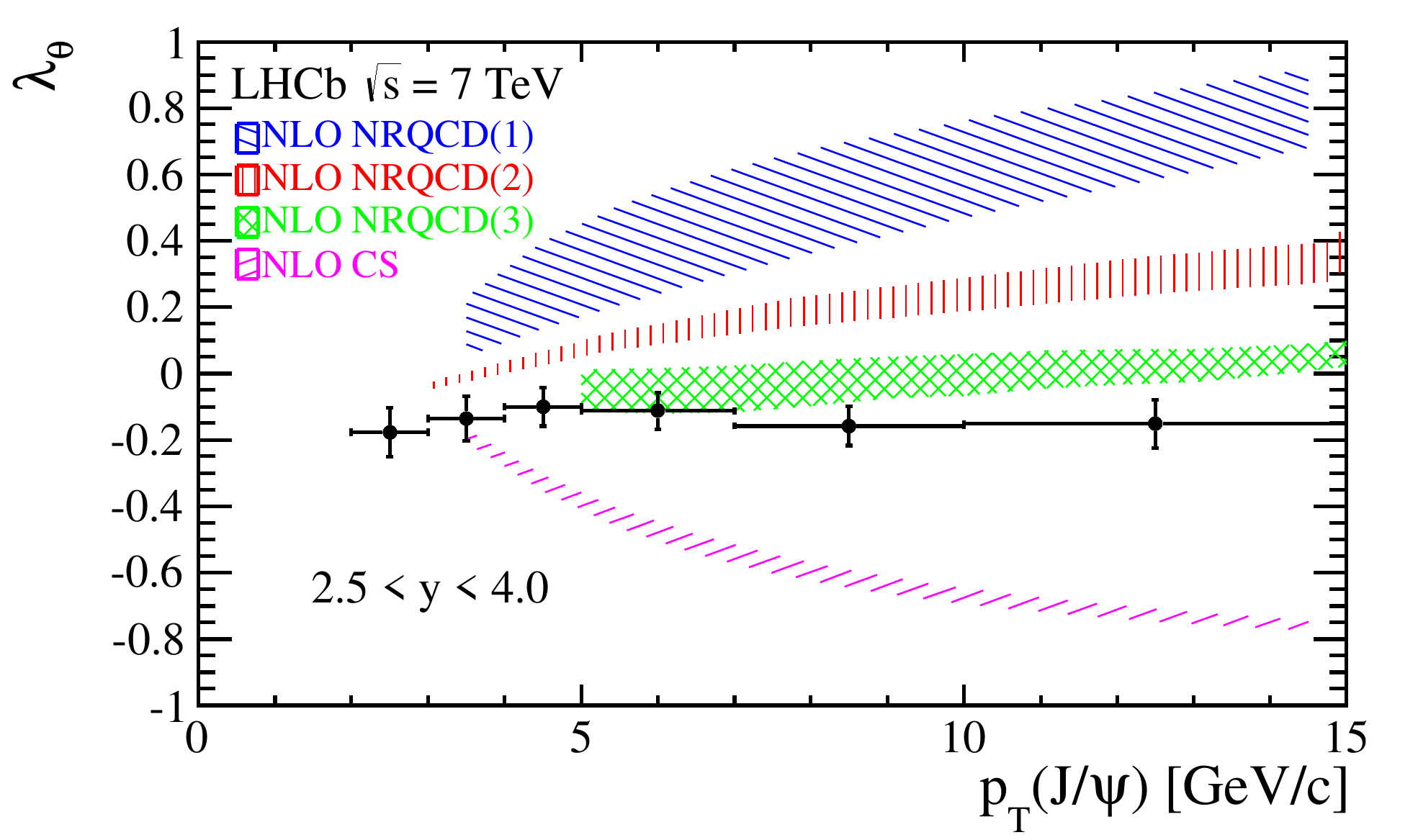} 
    \caption{Transverse momentum dependence of $J\!/\!\psi$ production cross-section ({\em left}) 
    and spin-alignment parameter $\lambda_\theta$ ({\em right}), compared to models. From references~\cite{Jpsi_old_ATLAS} and
 \cite{Jpsi_polarisation}.} 
    \label{fig:quarkonia-illustration}
  \end{center}
\end{figure}

The  {\em Colour Evaporation Model}  (CEM)~\cite{ref:CEM-1, ref:CEM-2} relates the production cross-section of quarkonia to the production of open heavy flavour, in the limit of zero relative momentum and with the invariant mass of the $Q\, \overline Q$ pair below the threshold for hadron pair production  
($2 \times m_Q < m_{Q\,\overline Q} < 2 \times m(H_Q)$). 
Figure~\ref{fig:quarkonia-illustration}-{\em left} shows fair agreement between CEM and observed differential cross-section for $J\!/\!\psi$~\cite{Jpsi_old_ATLAS}.

The {\em Colour Singlet Model} (CSM)~\cite{ref:CSM-2} assumes that a pertubative computation should be made including in the parton scattering only diagrams where the $Q\,\overline Q$ pair is produced in a colour-singlet configuration. The model has predictive powers since the only required external input is the amplitude of the wave function of the $Q\overline Q$ system at zero separation ($|\Psi(r\!=\!0)|^2$), 
which can be extracted from decay widths. 
NLO and partial next-to-next leading order (NNLO$^*$) computation~\cite{ref:CSM_NLO} have been found to modify significantly the predicted cross-section, as shown in Figure~\ref{fig:quarkonia-illustration}-{\em left}.

The {\em Non-Relativistic QCD} (NRQCD)~\cite{ref:NRQCD}  approach describes quarkonia production in terms of an effective theory, where long-distance effects are treated with non-perturbative matrix elements (LDME). 
In this scheme $Q\,\overline Q$ pairs produced in a coloured state ({\em colour-octet}) are included in the 
computation, assuming a transition into colourless physical states through the emission of soft gluons.
\begin{figure}[!b]
  \begin{center}
    \includegraphics[width=1\textwidth]{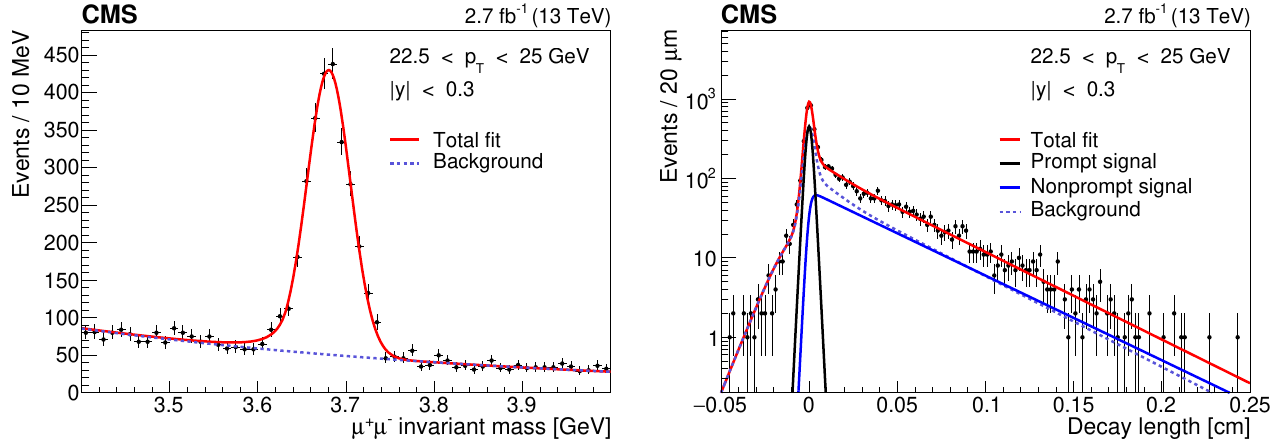} 
    \caption{Simultaneous fit to invariant mass and vertex separation for $\psi \mathrm{(2S)} \rightarrow \mu^+\mu^-$.
    From reference~\cite{QQ_CMS_13TeV}.}
    \label{fig:Psi(2s)-fit-CMS}
  \end{center}
\end{figure}
\begin{figure}[!t]
  \begin{center}
  \hspace{-0.7cm}
    \includegraphics[width=0.33\textwidth]{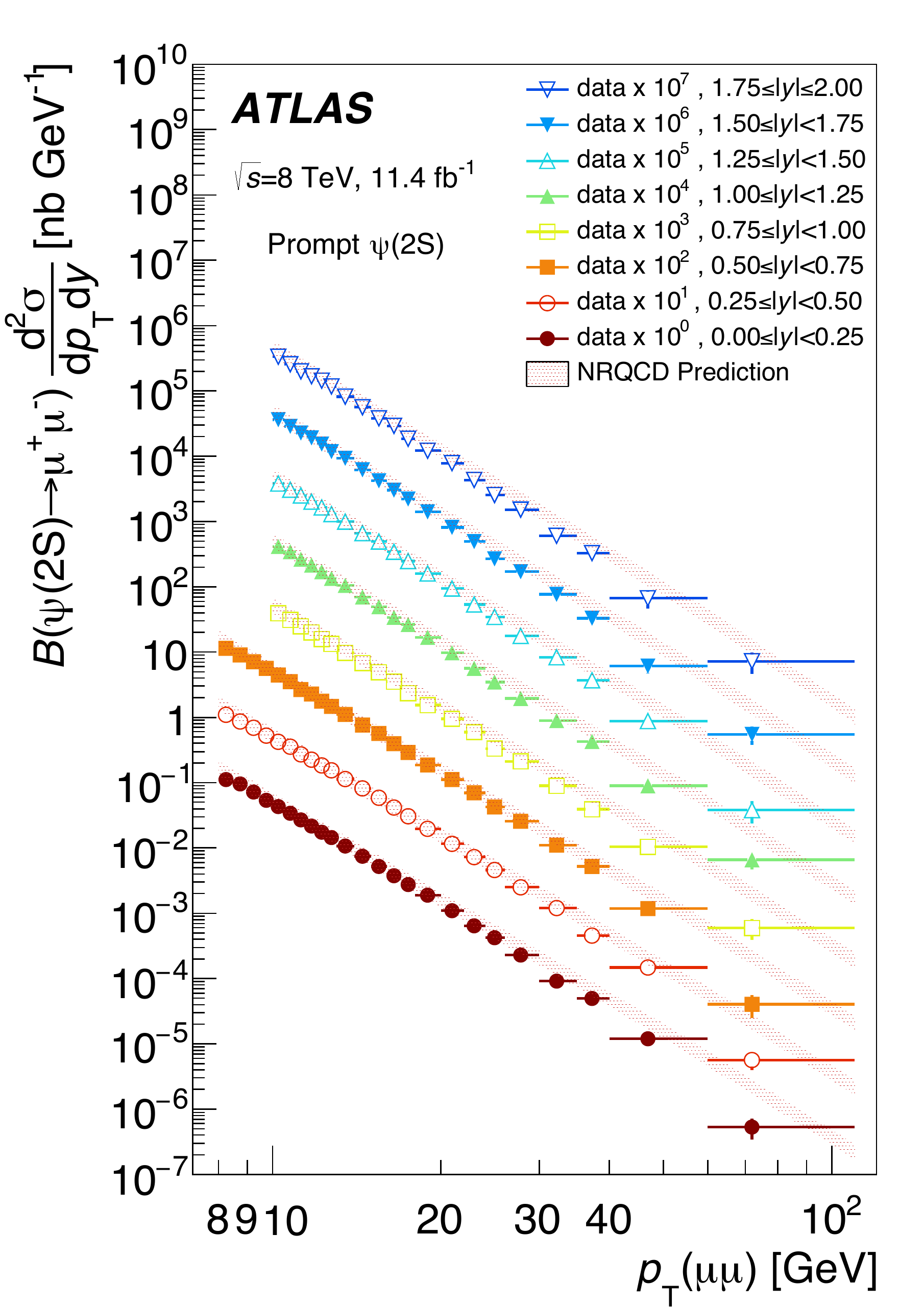} 
  \hspace{-0.25cm}
     \includegraphics[width=0.33\textwidth]{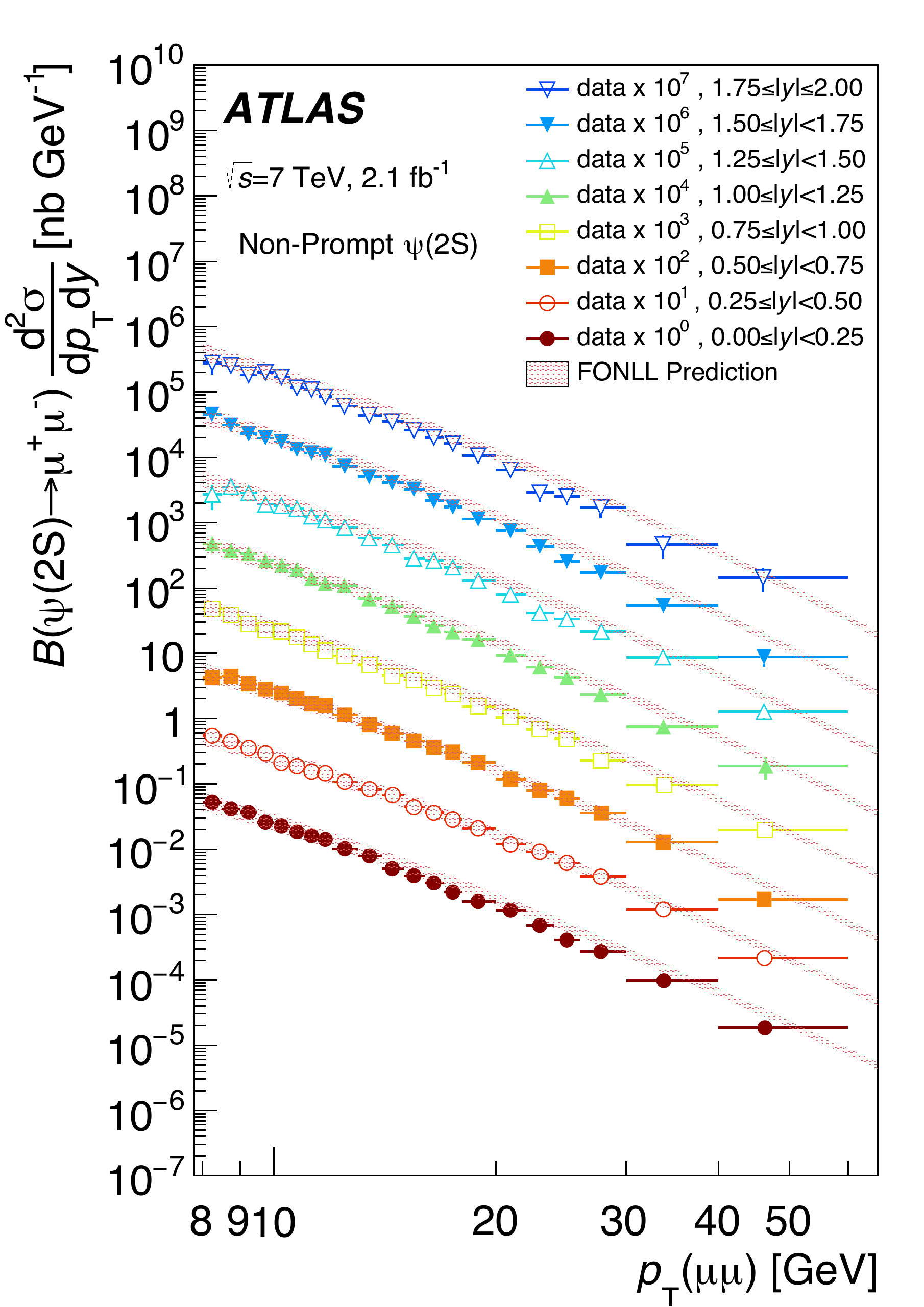} 
   \hspace{-0.23cm}
      \includegraphics[width=0.385\textwidth]{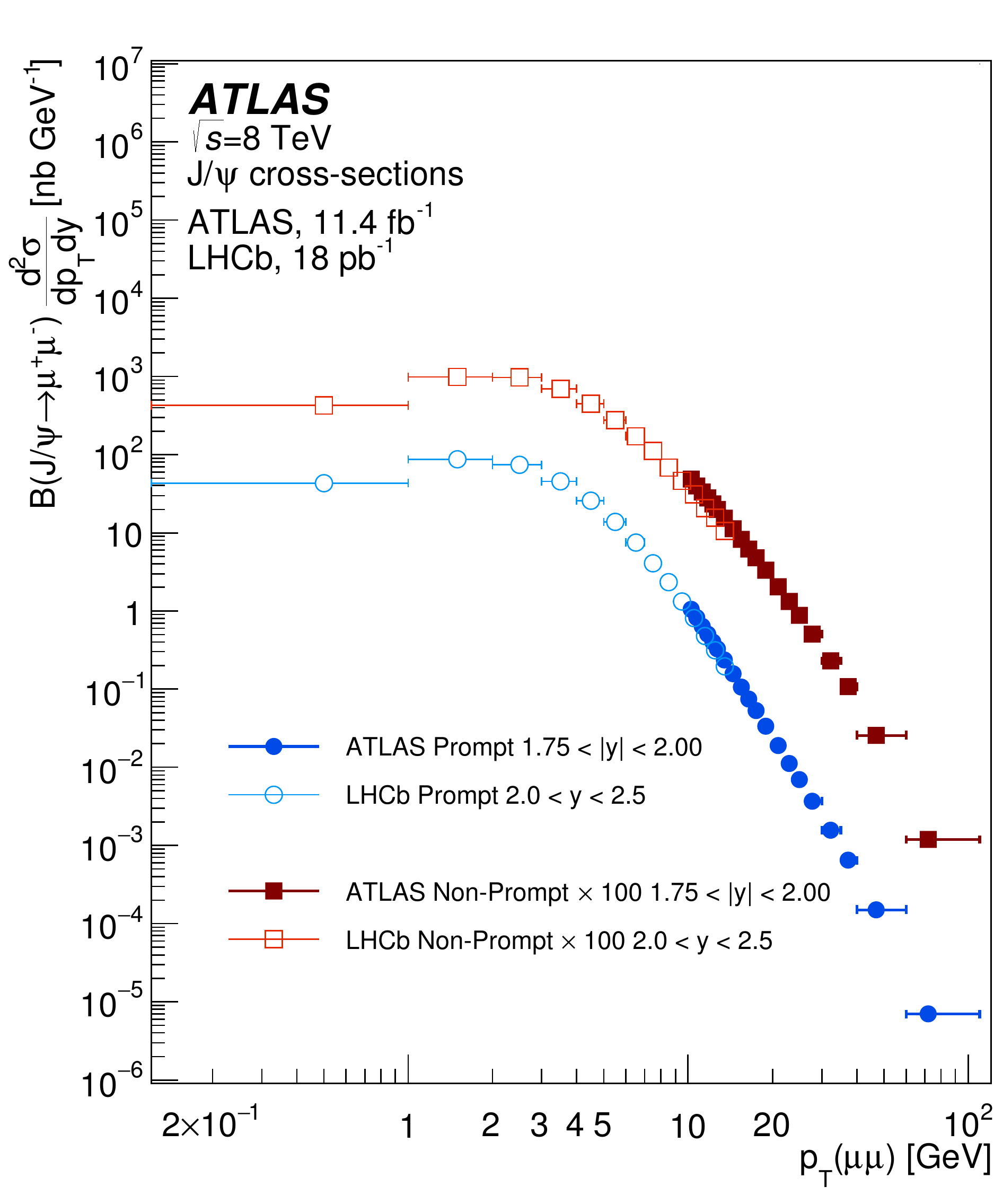} 
    \caption{$\psi$(2S) prompt and non prompt transverse momentum spectrum, 
    and $J\!/\!\psi$  spectra over a wide momentum range, from two measurements.
     From reference \cite{Jpsi_psi2S_ATLAS}.}
    \label{fig:JpsiATLAS_LHCb}
  \end{center}
\end{figure}
\subsection{Measurements of quarkonia production}
Measurements of quarkonia production have been performed at the LHC for all available collision energies. 
Detailed studies have been performed recently on vector states ($J\!/\!\psi$, $\psi$(2S), $\Upsilon$(nS), n = 1, 2, 3) decaying to muon pairs~\cite{Jpsi_Y_LHCb,Y_jpsi_psi2_CMS, Jpsi_psi2S_ATLAS, QQ_CMS_13TeV}.
 Measurements of $\psi$(2S) $\rightarrow J\!/\!\psi\,\pi^+\pi^-$~\cite{Jpsi_pi_pi_ATLAS}, 
 $\chi_c \rightarrow J\!/\!\psi\,\gamma$~\cite{Chi_c_ATLAS},
 $\chi_b \rightarrow \Upsilon \, \gamma$~\cite{Chi_b_LHCb_1_2, Chi_b_CMS}
 and $\eta_c$(1S) $\rightarrow p\,\overline p$~\cite{eta_c_LHCb} are also available.  
 For charmonium, the distinction between prompt and non-prompt production from $b$-hadron decays is obtained from a fit to the distance between the $p\,p$ interaction vertex and the muon-pair secondary vertex, which reflects the relatively long $b$-hadron lifetime and allows for an accurate separation.
This technique, which is also used for the selection of $b$-  and $c$-hadrons from background events, 
is illustrated in figure~\ref{fig:Psi(2s)-fit-CMS}.   Conversely, feed-down from higher mass quarkonia state ({\em e.g.}\ $\chi_{c\,(b)}$ to $J\!/\!\psi\;(\Upsilon)$ transitions) is usually not subtracted from the experimental data, but can be evaluated from cross-section measurements and known branching fractions.

Figure~\ref{fig:JpsiATLAS_LHCb} shows a measurement of the differential  $d^2\sigma\!/\!dp_{\mathrm T}dy$ production cross-section for prompt and non-prompt $\psi$(2S), 
and the comparison between nearby rapidity interval accessible in central (ATLAS in this case) and forward  (LHCb)
detectors. The measurements are in good agreement 
with the NRQCD prediction for prompt production, and FONLL for non-prompt production. 
Note the comparable magnitude and the different $p_{\mathrm T}$ dependence of the two production mechanism.
The colour-octet LDMEs used in the NRQCD prediction~\cite{NRQCD-Ma-Wang-Chao} are extracted from a fit to Tevatron data~\cite{CDF-Jpsi-psi2s}.  An apparent continuity in the production behaviour has been observed between 
$\sqrt s=2$~TeV studied at the Tevatron and 7, 8, 13~TeV at the LHC. 

\subsection{Polarisation of quarkonia}
The polarisation of quarkonia states has attracted attention because of the different predictions provided by the  production models. For vector states 
({\em e.g.}\ $J\!/\!\psi$, $\psi$(2S), $\Upsilon$(nS)) decaying into muon pairs
the angular distribution in the rest frame of the decaying particle is described by the parameters
$\lambda_\theta$, $\lambda_\phi$, $\lambda_{\theta\phi}$~\cite{Faccioli, Palestini}. 
For  $p_{\mathrm T} \gg m_{Q\overline Q}$ 
the CSM model predicts longitudinal spin-alignment, {\em i.e.}
$\lambda_\theta \simeq -1$, in the quarkonium helicity frame, while NRQCD favours transverse spin-alignment,
although different  level of polarisation have been predicted
($\lambda_\theta \simeq 0$ to $+1$, $\lambda_\phi \simeq 0$). 
The predictions are illustrated in figure~\ref{fig:quarkonia-illustration}-{\em right}, for 
$J\!/\!\psi \rightarrow \mu^+\mu^-$, together with a measurement~\cite{Jpsi_polarisation}. 

\begin{figure}[!tb]
\begin{center}
   \begin{minipage}{0.48\textwidth}
   \begin{center}
 \hspace{-0.8cm}
   \includegraphics[width=1.085\textwidth]{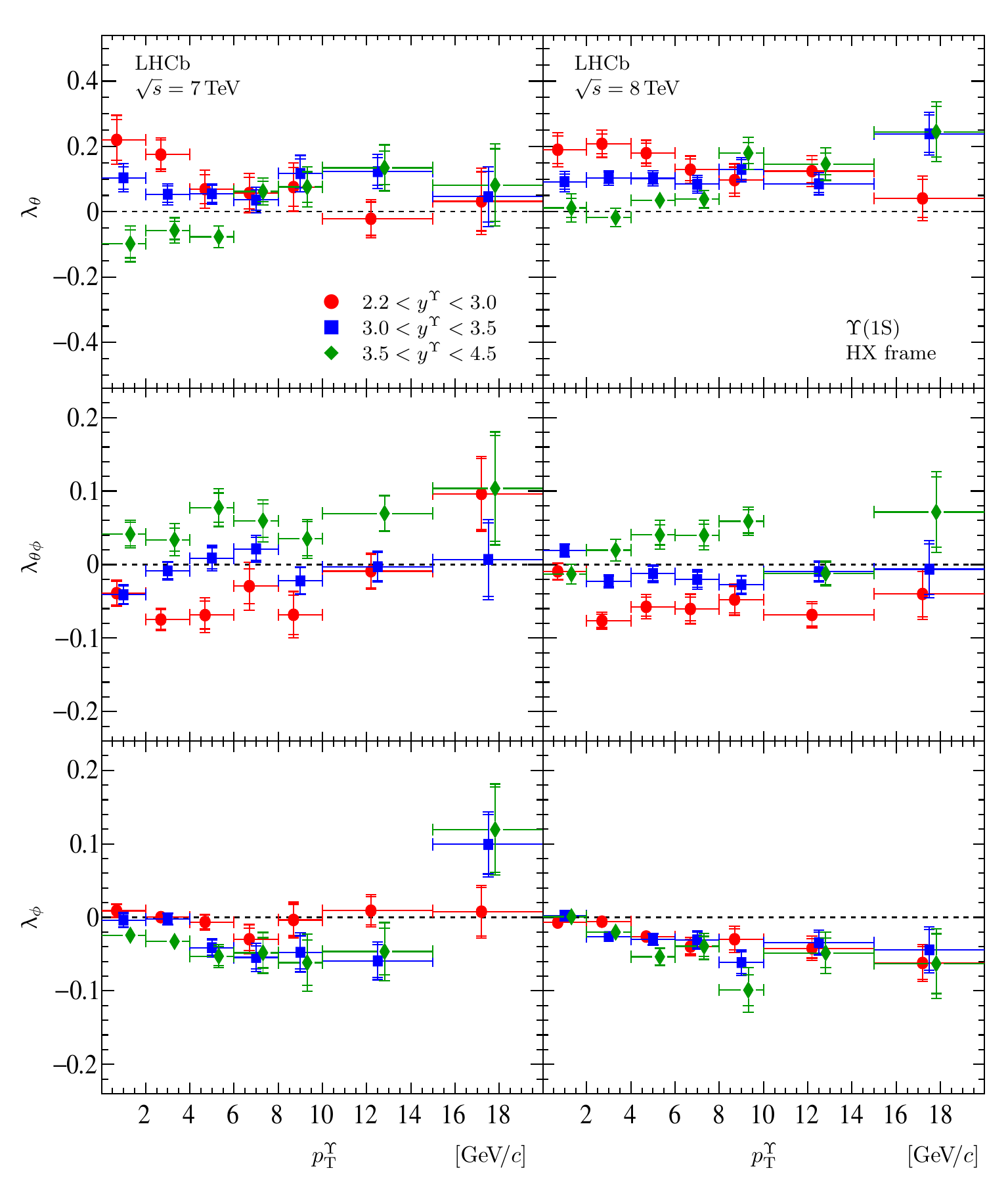} 
     \caption{$\Upsilon$(1S) polarisation: the three parameters describing the angular distribution for the decay to muon pairs are shown in $p_{\mathrm T}, y$ intervals, in the helicity frame. 
     From reference~\cite{Y_polarisation_LHCb}.} 
    \label{fig:Y-polarisation}
    \end{center}
    \end{minipage}
    \hspace{0.5cm}
    \begin{minipage}{0.46\textwidth}
    \begin{center}
        \includegraphics[width=0.78\textwidth]{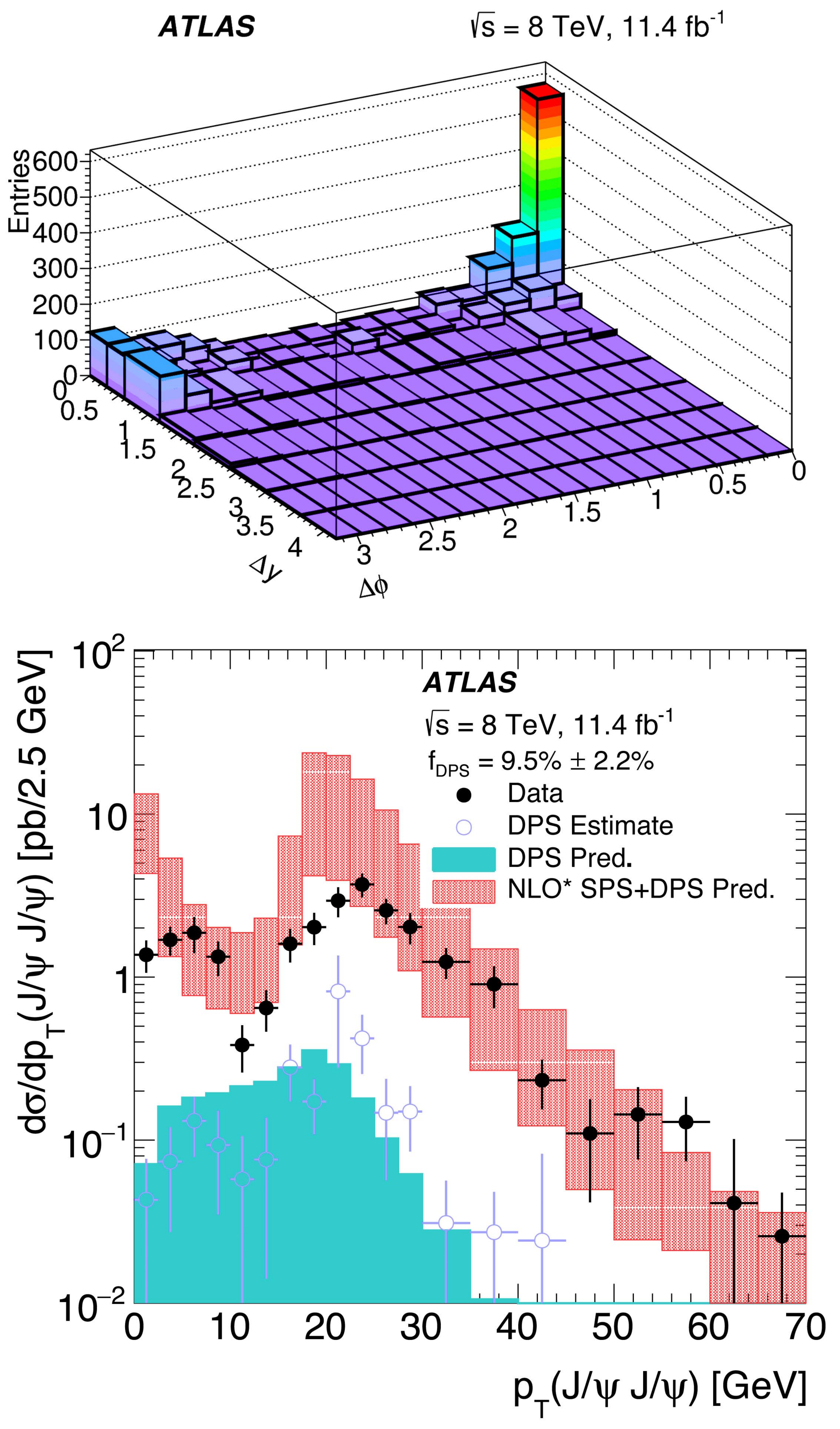} 
   \caption{Double prompt $J\!/\!\psi$ prompt production. {\em Top:} angular separation for SPS component.  
   {\em Bottom:} differential cross-section vs.\ $J\!/\!\psi\, J\!/\!\psi$ total transverse momentum.  
   From reference~\cite{2jpsi_ATLAS}.}
   \label{fig:double_Jpsi_ATLAS}
   \end{center}
   \end{minipage}
   \end{center}
   \end{figure}

Similar observations have been performed for $\psi$(2S) and 
$\Upsilon$(nS)~\cite{Y_Jpsi_psi2S_Ymult_polarisation_CMS, Y_polarisation_LHCb}. An example is shown in 
figure~\ref{fig:Y-polarisation}. 
The comparison with predictions requires the evaluation of the effects of feed-down from quarkonia states with higher mass.
In order to deal with the somewhat arbitrary choice of the reference frame, the data are usually analysed fitting for all parameters of the muons angular distribution, and also considering different choices of frames. 
Furthermore, combinations of angular distribution parameters that do not depend on the choice of frame have 
been identified~\cite{Faccioli, Palestini}.
So far, the measurements have not shown significant deviations form uniform angular distribution in either the high
$p_{\mathrm T}$ or the large $y$ regions. 

\subsection{Measurements of associated production}
In recent years the LHC experiments have performed several measurements of associated production, 
including  prompt $J\!/\!\psi$ together with $W$ or $Z$, double $J\!/\!\psi$, and, most recently, double $\Upsilon$
production, observed by CMS~\cite{ref:2Y_CMS}.
Recent measurements of double $J\!/\!\psi$ production are reviewed here. 
The interest in these studies lies both in the test of higher order QCD processes~\cite{Lansberg-2jpsi, Sun-Han-Chao} in the single parton scattering (SPS), and in the study of double parton scattering (DPS)~\cite{2jpsi_DPS}. 

LHCb~\cite{ref:2jpsi_LHCb} has performed a study where both $J\!/\!\psi$ are in the forward ($2 < y^{\,J\!/\!\psi} < 4.5$), low transverse momentum ($p_{\mathrm T}{}^{J\!/\!\psi} < 10$~GeV) region.  The observation is in agreement with the expectations, but does not allow a model independent separation of the SPS and DPS components. 

The situation is more favorable in the $|y^{\,J\!/\!\psi}|<2.1$, $p_{\mathrm T}{}^{J\!/\!\psi} > 8.5$~GeV region, 
where ATLAS~\cite{2jpsi_ATLAS}
has used both $\Delta \phi$ and the rapidity separation $\Delta y$ to determine the uncorrelated DPS component. 
Figure~\ref{fig:double_Jpsi_ATLAS} shows the angular separation for the SPS component, 
peaking at $\Delta y \simeq 0$ and both $\Delta \phi \simeq 0$ and $\simeq \pi$. 
Also shown is is the production cross-section as a function of the transverse momentum of the pair, 
which presents as well a bi-modal distribution 
with a peak at $p_{\mathrm T}{}^{J\!/\!\psi\, J\!/\!\psi} \approx 25$~GeV, 
in fair agreement with NLO SPS computation.   
From the measurement of the DPS contribution $\sigma_\mathrm{DPS}^{J\!/\!\psi\, J\!/\!\psi} = 14.8 \pm 3.8$~pb 
and of the prompt production cross-section $\sigma^{\,J\!/\!\psi}\!$, the effective cross-section for DPS is derived as
$\sigma_\mathrm {eff}^{J\!/\!\psi\, J\!/\!\psi} \equiv \left( \sigma^{\,J\!/\!\psi}\right ){}^2 / \left( 2 \times \sigma_\mathrm {DPS}^{\,J\!/\!\psi\, J\!/\!\psi}\right )\,$, and found equal to $6.3 \pm 1.9 \; \mathrm {mb},$ lower than the values $\approx \!15$~mb
observed for most other DPS processes~\cite{2jpsi_ATLAS}.

\section{Conclusions, acknowledgments} 
The production of heavy quarks hadrons and quarkonia has been a very active area of studies in recent years, 
thanks mainly to measurements performed at the LHC and to theory developments in the areas of pertubative and non-relativistic QCD. Correlation in $b$--$\overline b$ production and associated quarkonia production 
are recently added observables, and others are likely to join in the near fu\-ture, as new data are being collected.

Sincere thanks to the organisers of {\em HADRON-2017} for the invitation to present this subject, and to
Leonid Gladilin, Vakhtang Kartvelishvili, and James Walder for useful discussions.

\end{document}